\documentclass[sigconf]{acmart}
\usepackage{subfigure}
\usepackage{multirow}
\usepackage{tabularx}
\usepackage{adjustbox}
\usepackage{tikz}

\AtBeginDocument{%
  \providecommand\BibTeX{{%
    \normalfont B\kern-0.5em{\scshape i\kern-0.25em b}\kern-0.8em\TeX}}}


\copyrightyear{2021}
\acmYear{2021}
\setcopyright{acmlicensed}\acmConference[CIKM '21]{Proceedings of the 30th ACM International Conference on Information and Knowledge Management}{November 1--5, 2021}{Virtual Event, QLD, Australia}
\acmBooktitle{Proceedings of the 30th ACM International Conference on Information and Knowledge Management (CIKM '21), November 1--5, 2021, Virtual Event, QLD, Australia}
\acmPrice{15.00}
\acmDOI{10.1145/3459637.3482065}
\acmISBN{978-1-4503-8446-9/21/11}

\begin{document}
\fancyhead{}
\title{Binary Code based Hash Embedding for Web-scale Applications}

\author{
 Bencheng Yan$^{*}$, Pengjie Wang$^{*}$, Jinquan Liu, Wei Lin, Kuang-Chih Lee, Jian Xu and Bo Zheng$^{\dagger}$
 }
  \affiliation{%
  \institution{Alibaba Group}
}
 \email{{bencheng.ybc,pengjie.wpj,vjinquan.ljq,kuang-chih.lee,xiyu.xj,bozheng}@alibaba-inc.com, lwsaviola@163.com}
 \thanks{$*$ These authors contributed equally to this work and are co-first authors.}
 \thanks{$\dagger$ Corresponding author}

\begin{abstract}
Nowadays, deep learning models are widely adopted in web-scale applications such as recommender systems, and online advertising. 
In these applications, embedding learning of categorical features is crucial to the success of deep learning models. 
In these models, a standard method is that each categorical feature value is assigned a unique embedding vector which can be learned and optimized. 
Although this method can well capture the characteristics of the categorical features and promise good performance, it can incur a huge memory cost to store the embedding table, especially for those web-scale applications. 
Such a huge memory cost significantly holds back the effectiveness and usability of EDRMs. 
In this paper, we propose a binary code based hash embedding method which allows the size of the embedding table to be reduced in arbitrary scale without compromising too much performance.
Experimental evaluation results show that one can still achieve 99\% performance even if the embedding table size is reduced 1000$\times$ smaller than the original one with our proposed method.
\end{abstract}



\vspace{-1em}

\maketitle

\section{Introduction}
\label{sec:Introduction}
\begin{figure}[t]
\centering
\includegraphics[width = .4\textwidth]{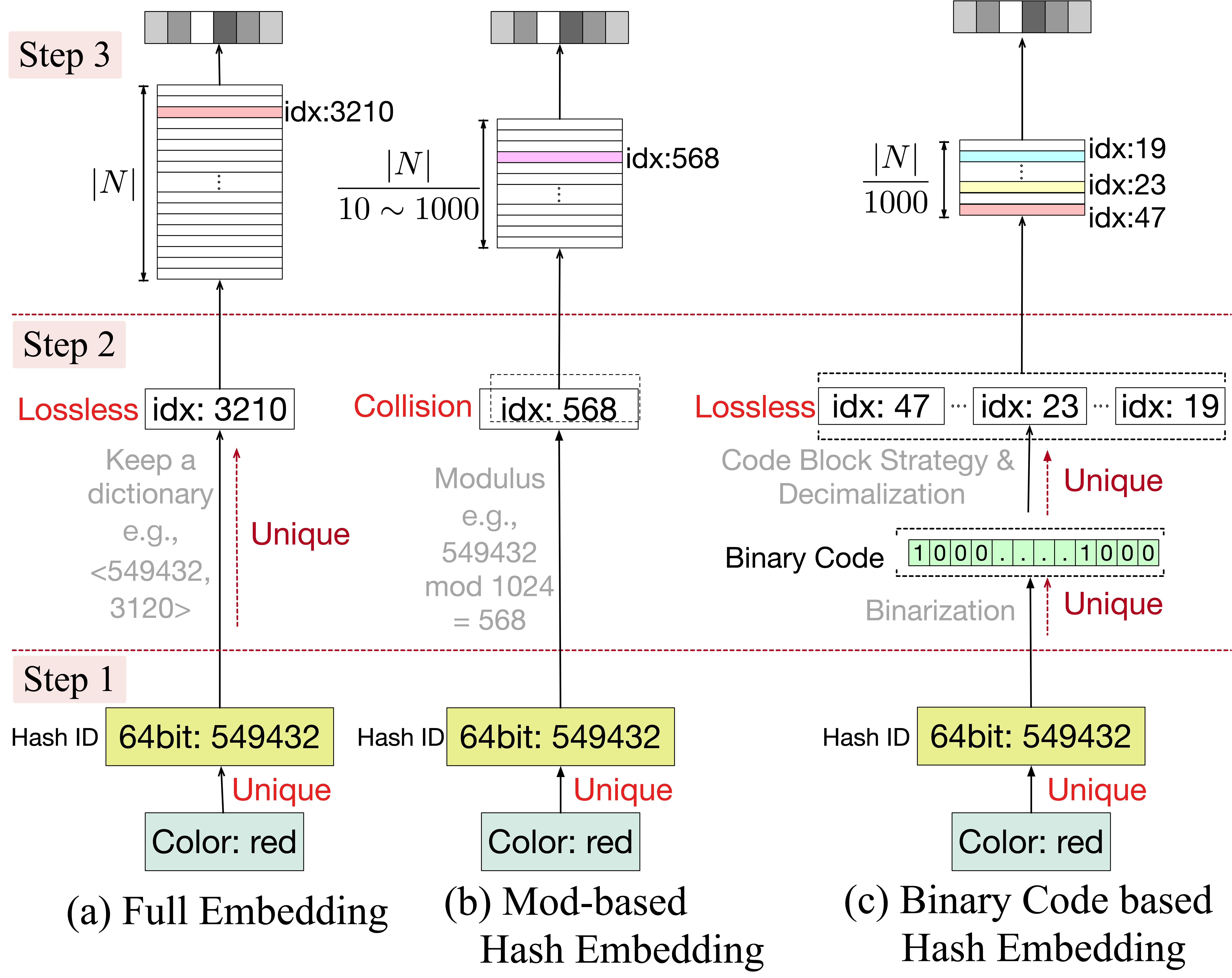}
\vspace{-1em}
\caption{
Comparisons of different embedding methods.
 Step 1, 2, and 3 refer to feature hashing, embedding index generation, and embedding generation respectively.
}
\vspace{-1.5em}
\label{figure:hash_compare}
\end{figure}

Embedding learning for categorical features plays an important role in embedding-based deep recommendation models (EDRMs) \cite{cheng2016wide,guo2017deepfm,wang2017deep}.
A standard method, often referred to as \emph{full embedding}, for embedding learning is to learn the representation of each feature value \cite{shi2020compositional}. 
Specifically, let $F$ be a categorical feature and $|F|$ be its vocabulary size, each feature value $f_i \in F$ is assigned an embedding index $k_i$ so that the $k_i$-th row of the embedding table $W\in \mathbb{R}^{N \times D}$ is the embedding vector of $f_i$, where $N=|F|$ in the full embedding method and $D$ is the embedding dimensionality (see Fig \ref{figure:hash_compare} (a)).

However, such full embedding learning suffers from severe memory cost problems. 
Actually, the memory cost of the embedding table is $O(|F|D)$ which grows linearly with $|F|$. 
For web-scale applications, one may need to store a huge embedding table since the vocabulary size may be millions or even billions.
For example, suppose $|F| =$ 500 million and $D = 256$, the corresponding memory cost will be 475GB. 
In practice, such a huge cost becomes a bottleneck in deploying EDRMs in memory-sensitive scenarios.

Therefore, it is crucial to reduce the size of the embedding table \cite{shi2020compositional,chen2018learning}. 
In this paper, we highlight two challenges:
(1) \emph{\textbf{Challenge one: Flexibility}}. 
The memory constraint varies with different scenarios (from distributed servers to mobile devices).
The embedding reduction methods need to be flexible enough to meet different memory requirements. 
Especially for mobile devices, a tiny EDRM is needed to meet the limited memory requirement.
(2) \emph{\textbf{Challenge two: Performance Retention}}.
Since a big model usually has a better capacity and hence a better performance, embedding reduction may bring a performance gap due to the fewer parameters used in the reduced model.
Hence, how to keep high performance when the memory size is reduced is a big challenge, especially for the memory-sensitive scenarios (e.g., in mobile devices).

In general, there are two directions to reduce the embedding table size, i.e., reducing the size of each embedding vector and reducing the number (i.e., $N$) of the embedding vectors in an embedding table. 
The embedding table size of the former methods (e.g., product quantization \cite{jegou2010product,ge2013optimized}, K-D method \cite{chen2018learning,kang2020learning,li2019multi,khrulkov2019tensorized,shu2017compressing}, and AutoDim \cite{liu2021learnable,liu2020automated,zhao2020autoemb,ginart2019mixed,joglekar2020neural,zhao2020memory}) is still linearly increased with $|F|$, failing to tackle the memory problem caused by a large vocabulary size in web-scale applications \cite{shi2020compositional}.
Hence these methods are not considered in our paper. 
For the latter methods, 
they typically apply a \emph{mod-based hash embedding} to reduce $N$. 
The key idea of them is to apply modulo operation on the unique Hash ID of each feature value, i.e., focusing on Step 2 in Fig \ref{figure:hash_compare} (b).
For example,  Hash embedding \cite{weinberger2009feature} takes the remainder of the Hash ID divided by $M$ as the embedding index, reducing the embedding size from $O(|F|D)$ to $O(MD)$.
 The problem of this method is that different feature values may have the same embedding index and hence the same embedding vector, leading to poor performance.
Multi-Hash (MH) \cite{tito2017hash} 
 adopts multiple embedding indices for one feature value,  reducing the collision rate among feature values.
But different feature values may still be indistinguishable especially for a tiny model, failing to the challenge two.
Q-R trick \cite{shi2020compositional} uses both the remainder and the quotient as embedding indices to identify a feature value.
However, Q-R trick fails to the challenge one since its minimal reduced size is related to $\sqrt{|F|}$ rather than any scales.
Although the generalized Q-R tries to address this problem, it needs a lot of effort to design the divisor \cite{shi2020compositional}.
The comparisons are summarized in Table \ref{table:Comparison about embedding methods on addressing the two challenges (variety and performance)}.

In this paper, 
unlike the existing methods which adopt a modulo (collision) operation, we bring the idea of binary code (e.g., the binary code of integer 13 is $1101_2$) which is unique for different Hash ID and propose a \emph{binary code based hash embedding} method to tackle this reduction problem (see Fig \ref{figure:hash_compare} (c)).
Specifically, we first binarize the Hash ID into a binary code.
Then, to address the challenge one, we propose a code block strategy and reduce the embedding table size by adjusting the code block length flexibly.
To address the challenge two, the generated embedding index is designed to be unique for different feature values at any reduction ratios.
The uniqueness at any reducing ratios allows EDRMs to distinguish different feature values, 
leading to a good performance even for a tiny model.
Furthermore, Step 2 of our method is a deterministic and non-parametric process and can be computed on-the-fly.
This property is friendly for EDRMs both on the convenient application and handling new (out-of-vocabulary) feature values.


We also note that we are aware of some recent works using similar terms such as \emph{learning binary embedding} \cite{hong2017fried,kulis2009learning,yi2015binary}. 
We want to point out that they are in totally different contexts. 
In these works, binary refers to that each element in an embedding vector is a binary number for a fast similar embedding search.
While in our work, binary refers to binarize the integer ID into a binary code.

To summarize, the main contributions are listed as follows:
(1) We propose binary code based hash embedding, a simple but effective embedding method, to reduce the embedding table size and keep a high performance at the same time.
(2) A code block strategy is presented to adjust the embedding table size flexibly and a lossless embedding index generation process is elaborately designed to allow the model to distinguish different feature values and achieve better performance.
(3) Experimental results on large-scale real-world datasets show that with the help of the proposed method, the model size can be 1000$\times$ smaller than the original model, and keep 99\% performance as the original model achieves at the same time.



\begin{table}[]\smaller
\caption{Comparison about embedding methods.
 }
 \vspace{-1em}
\begin{tabular}{ccccccc}
\toprule 
            & Full & Hash & MH & Q-R & Ours\\ \hline
Flexibility     &  Bad  &    Good  &     Good       &    Fair&   Good             \\ 
Performance Retention&   Good   &   Bad   &      Fair    &    Good &   Good            \\ \bottomrule
\end{tabular}
 \vspace{-2em}
\label{table:Comparison about embedding methods on addressing the two challenges (variety and performance)}
\end{table}

\section{Binary Code based Hash Embedding}
In this section, we introduce the framework (see Fig \ref{figure:framework}) of our methods.
In general, we also adopt three steps as introduced in Fig \ref{figure:hash_compare}.

\subsection{Feature Hashing}
\label{sec:Feature Hash}

\begin{figure}[t]
\centering
\includegraphics[width = .4\textwidth]{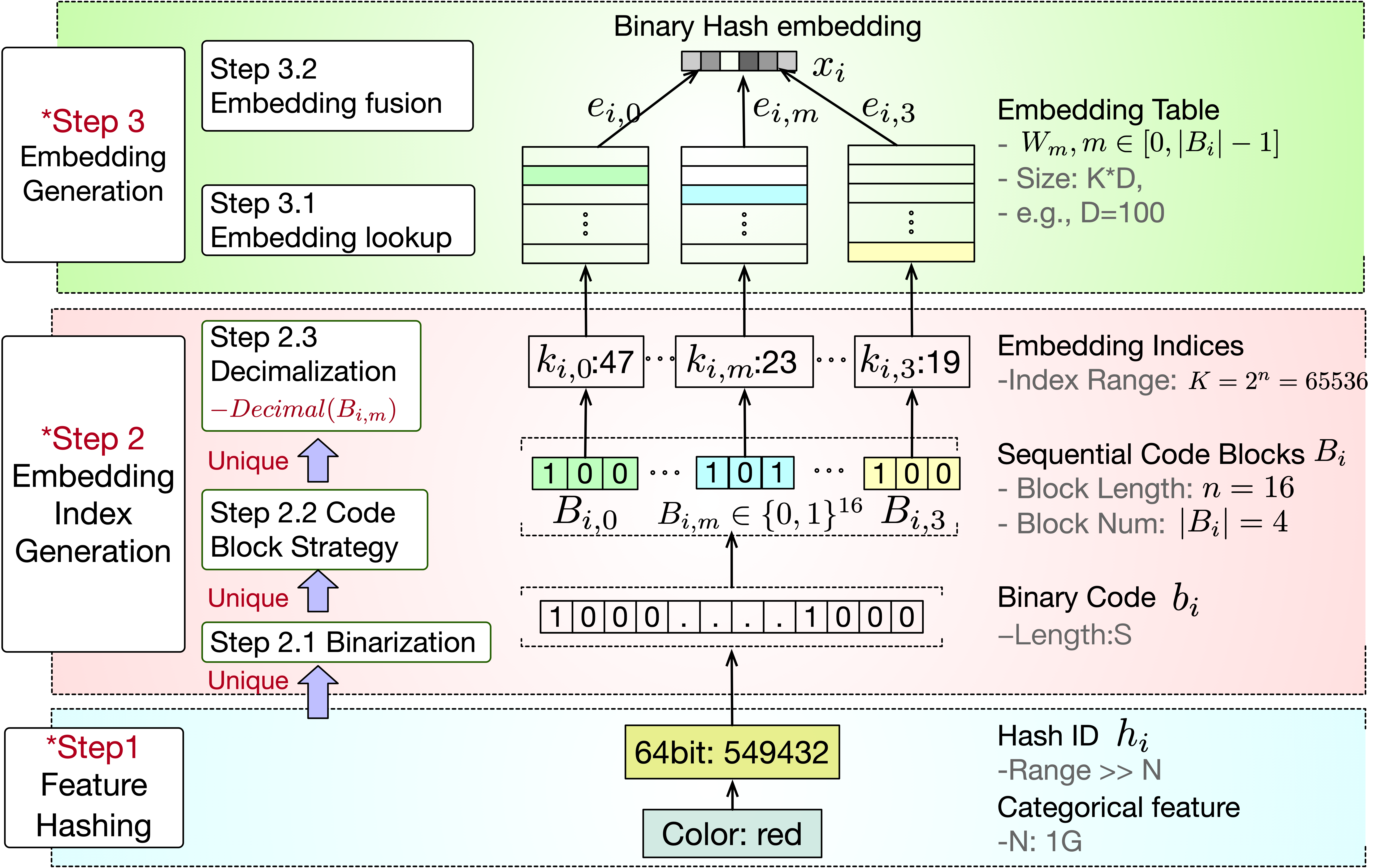}
\vspace{-1em}
\caption{The framework of the proposed method.}
\vspace{-2em}
\label{figure:framework}
\end{figure}

In practice, the raw categorical feature values may be represented as various types, such as String and Integer values. 
To handle different types of categorical feature values, in practice, a feature hashing \cite{tito2017hash,shi2020compositional,weinberger2009feature} is firstly applied to map these raw feature values into a uniformed integer number, called Hash ID (see Fig \ref{figure:framework}).
Formally, the feature hashing process can be expressed as
$
h_i=\mathcal{H}(f_i)
$
where $\mathcal{H}$ refers to a hash function (e.g., Murmur Hash \cite{yamaguchi2013hardware}) and $h_i$ is an integer number, called the Hash ID of $f_i$.
In practice, the output length of $\mathcal{H}$ is always a large value (e.g., $h_i$ is a 64-bits integer) to make the collision among $h_i$ as small as possible.
In this case, $h_i$ can be basically taken as a unique ID for different $f_i$ \cite{weinberger2009feature,kang2020deep}.

\subsection{Embedding Index Generation}
In this section, we introduce the embedding index generation process including binarization, code block strategy, and decimalization.

\subsubsection{Binarization}
\label{sec:Binary Code Generation}
After Step 1, each feature value $f_i$ is mapped to $h_i$, which is basically regarded as a non-collision mapping due to the large output space \cite{kang2020deep,weinberger2009feature}.
Then the binary code $b_i \in \{0,1\}^S$ (where $S$ refers the binary code length) of $f_i$ can be generated by transforming this unique $h_i$ to a binary form (e.g., the binary code of integer 13 is $1101_2$).
Note $b_i$ is also unique for different $f_i$.

\subsubsection{Code Block Strategy}
\label{sec:Code Block Strategy}
To allow the model can flexibly reduce memory, we propose a novel strategy called code block strategy.
Generally speaking, the code block strategy divides each 0-1 value in $b_i$ to different blocks.
Then, the ordered 0-1 values (i.e., 0-1 code) in each block can represent $K=2^n$ unique integers where $n$ is the number of 0-1 values in this block (see Step 2.2 in Fig \ref{figure:framework}).

If we take the decimal form of 0-1 code in each block as an embedding index and map each index to an embedding table $W \in \mathbb{R}^{K\times D}$, the size of the embedding table can be flexibly adjusted by $n$.
For example, when $n=1$ (i.e., the number of 0-1 values in each block is 2), the embedding table size is $O(2D)$.
When all 0-1 values in $b_i$ are arranged into one block, the embedding table size is $O(|F|D)$ (i.e., full embedding).
In other words, by controlling the value of $n$, we can adjust the embedding table size to meet various scenarios (from distributed services to mobile devices).

Formally, we define $B_i=[B_{i,0};B_{i,1};...;B_{i,m};...]$ as the sequential code blocks produced by a code block strategy on $b_i$, and $|B_i|$ refers to the number of blocks.
Then the $m$-th code block $B_{i,m} \in \{0,1\}^n$ can be represented as 
\begin{align}
B_{i,m}=\textit{Order}(\{b_{i,j}|Alloc(b_{i,j})=m\})
\end{align}
where the function $\textit{Alloc}$ is the allocation function which allocates each 0-1 value to different blocks. 
$\textit{Order}$ is a function which gives an order for the 0-1 value in each block and generates a 0-1 code for each code block.
Here we give two code block strategies (including Succession and Skip) as examples to show how it works (other possible strategies can also be allowed).

\begin{figure}[t]
\centering
\includegraphics[width = .35\textwidth]{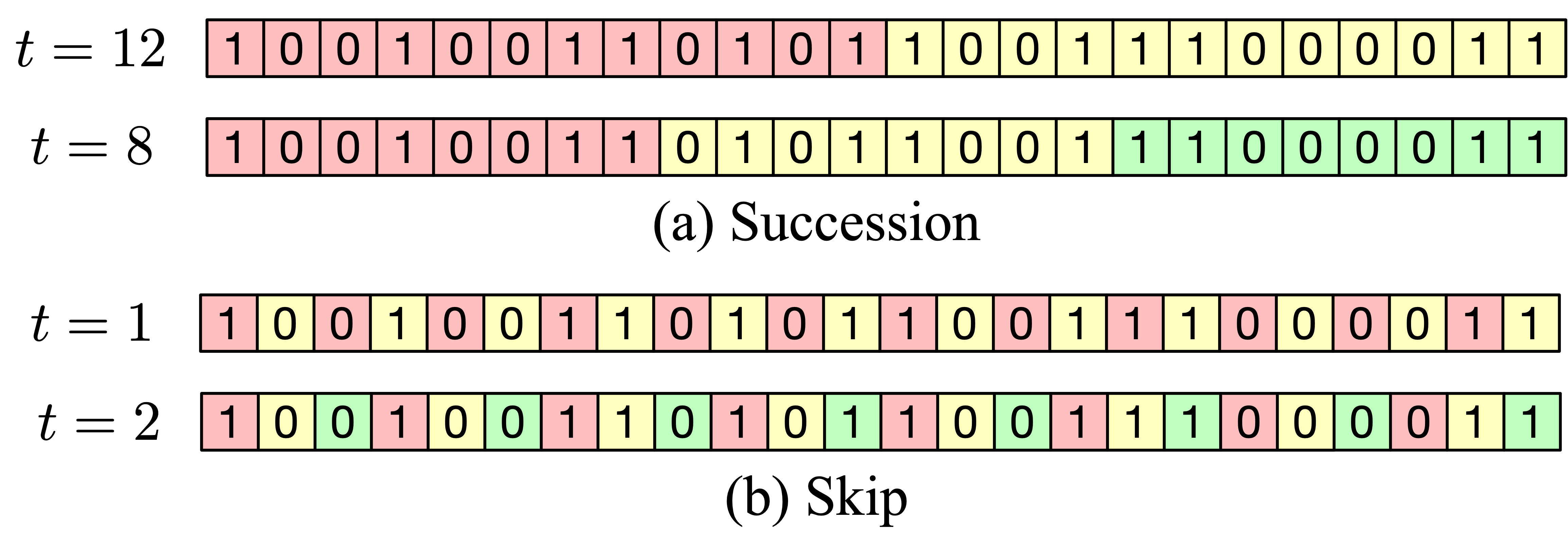}
\vspace{-1em}
\caption{Code block strategy examples. 
The binary code $b_i=100100110101100111000011$. 
The 0-1 values plotted with the same color in each case are divided into the same block.
}
\vspace{-1.5em}
\label{figure:code_block_strategy_example}
\end{figure}

\noindent\textbf{Succession.} As shown in Fig \ref{figure:code_block_strategy_example} (a), the succession strategy puts the $t$ successive 0-1 values in a binary code into the same block.
The $\textit{Order}$ function keeps 0-1 values in the same relative position in $b_i$.
Note if the number of the last 0-1 values in $b_i$ is less than $t$, all of the left values are divided into a new code block.

\noindent\textbf{Skip.} As shown in Fig \ref{figure:code_block_strategy_example} (b), if the number of interval values of two 0-1 values in a binary code is $t$, they will be divided into the same block.
The $\textit{Order}$ function is the same as that in Succession.

Note given one of the above code block strategies, we can obtain a unique sequence of code blocks $B_i$ for the binary code $b_i$.
This property guarantees the process of code block strategy is lossless.
  



\subsubsection{Decimalization}
\label{sec:Decimalization}
The embedding index of each block can be obtained by decimalizing $B_{i,m}$ (e.g., $Decimalize(1101_2) = 13$), i.e., 
$k_{i,m}=Decimalize(B_{i,m})$
where $k_{i,m}$ is the embedding index of $B_{i,m}$.

\subsection{Embedding Generation}
When obtaining multiple indices for $f_i$, to get its embedding, two steps are proposed, i.e., embedding lookup and embedding fusion.

\subsubsection{Embedding Lookup}
\label{sec:embedidng lookup sharing}
As introduced above, each code block $B_{i,m}$ in $B_i$ can obtain an embedding index $k_{i,m}$.
The number of blocks is $|B_i|$, leading to a total of $|B_i|$ embedding indices.
Then we can map each embedding index into an embedding vector, i.e., 
$
e_{i,m}=\mathcal{E}(W_{m},k_{i,m})
$
where $W_{m}$ is a embedding table,  $e_{i,m}$ refers to the embedding of $B_{i,m}$ and $\mathcal{E}$ is a embedding lookup function which usually returns the $k_{i,m}$-th row of $W_{m}$.
In practice, keeping $|B_i|$ embedding tables for different $B_{i,m}$ may also cost a lot memory consumption.
Therefore, it is common to keep a single embedding table and share this table among all $B_{i,m}$ \cite{tito2017hash}.



\subsubsection{Embedding Fusion}
\label{sec:Embedding Combination}
To generate the final embedding vector $x_i$ of $f_i$, an embedding fusion function $g$ is applied, 
\begin{align}
x_i=g(e_{i,0},e_{i,1},e_{i,2},...,e_{i,|B_i|-1})
\end{align}
The design of the fusion function can be various, such as pooling, LSTM, concatenation and so on.
In this paper, by default, we adopt sum pooling as the fusion function (others can also considered).

\begin{table*}[t] \smaller
\caption{The results for CTR tasks.
}
\vspace{-1em}
\begin{tabular}{c|ccccc|ccccc|ccccc}
\toprule 
Dataset & \multicolumn{5}{c|}{Alibaba}   & \multicolumn{5}{c|}{Amazon}& \multicolumn{5}{c}{MovieLens}                                      \\ \hline
Reduction Ratio                             & 0.1\%          & 0.75\%         & 1.5\%          & 5\%            & 37.5\% & 0.1\%          & 0.75\%         & 1.5\%          & 5\%            & 37.5\% & 0.1\%          & 0.75\%         & 1.5\%          & 5\%            & 37.5\%         \\ \hline 
Full                 & 70.57          & 70.57          & 70.57          & 70.57          & 70.57         & 68.56& 68.56& 68.56& 68.56& 68.56      & 80.23 &80.23 &80.23& 80.23& 80.23     \\ \hline
Hash                & 69.06          & 69.35          & 69.63          & 69.86          & 70.24         & 64.66 & 66.27& 66.66& 67.32 &67.67   & 73.62 &75.50  &  76.42& 77.68 &79.12   \\ 
MH         & 69.45          & 69.66          & 69.73          & 70.00          & 70.28          & 66.15 & 67.53& 67.58& 67.79 &68.03          & 74.97	&	78.06	&78.23&		78.85&		79.85       \\ 
Q-R                     & 69.47          & 69.58          & 69.82          & 70.10          & 70.28           & 66.67 & 67.43& 67.62& 67.73 &68.17    & 75.92		&78.05	&78.30		&78.90		&79.78   \\ \hline 
BH                  & \textbf{69.90} & \textbf{69.95} & \textbf{70.02} & \textbf{70.26} & \textbf{70.38} & \textbf{67.59} & \textbf{67.73}& \textbf{67.83}& \textbf{67.92} &\textbf{68.38} &\textbf{78.20}		&\textbf{78.38}	&\textbf{78.54}	&\textbf{79.14}		&\textbf{80.02}\\ \bottomrule
\end{tabular}
\label{table:The AUC results for CTR tasks on Alibaba Dataset.}
\vspace{-1em}
\end{table*}

\begin{table}[t] \smaller
\caption{
The results of memory size when all the methods archive 99 \% performance as the full embedding method achieves in AUC score.
}
\vspace{-1em}
\begin{adjustbox}{max width=\linewidth}
\begin{tabular}{c|cccccc}
\toprule 
      \multicolumn{2}{c}{}                         & Full& Hash  & Q-R & MH & BH \\ \hline
\multicolumn{2}{c}{Model size (G)}                  & 843.2          & 44.1           &  13.0         & 12.5                 & 0.8         \\ \hline
                                                             
\multirow{3}{*}{Top 3} &User ID                         & 288.94         & 10.16          &   3.11        & 2.88                 & 0.15        \\ 
&Item ID                         & 242.71         & 11.07          &    3.01       & 2.95                 & 0.16        \\ 
&Query ID                        & 165.24         & 11.07          &     3.17      & 3.15                 & 0.16        \\ 
\bottomrule
\end{tabular}
\end{adjustbox}
\vspace{-2em}
\label{table:The results of memory size when all the methods }
\end{table}

\subsection{Discussion}

\subsubsection{Desiderata}
\label{sec:Desiderata}
There are several key desiderata of our method, which EDRMs can be benefited.
(1) \textbf{Determinacy.}
The indices generation is a deterministic and non-parametric process.
It is computed on the fly,
making it simple to practical implementations and friendly to new feature values.
(2) \textbf{Flexibility.}
The size of embedding table $W \in \mathbb{R}^{K\times D}$ is mainly determined by $n$ (i.e., the number of 0-1 values in each code block).
It means the memory reduction ratio can be flexibly adjusted from $2/|F|$ to 1 (assuming adopting embedding table sharing strategy).
This benefits EDRMs can be developed on memory insensitive scenarios to sensitive scenarios.
(3) \textbf{Uniqueness.}
No matter what the reduction ratio is, $B_i$ is unique for each feature value.
This enables the model to distinguish different feature values and further improve the model performance.

\subsubsection{Sub-collision Problem}
We should point out although $B_i$ is unique, there may exist sub-collision among two feature values (e.g., $B_i \neq B_j$ but $B_{i,1}=B_{j,1}$), called sub-collision problem.
Actually, it is an open problem which exists in most mod-based hash methods \cite{weinberger2009feature,tito2017hash,shi2020compositional}.
We leave it as one of the future work.
In practice, a hash function (e.g., Murmur hash \cite{yamaguchi2013hardware}) which can randomly map values to a large space is used to relieve this problem.

\subsubsection{The Relation with Existing Methods}
\label{sec:The Relation between Binary Hash Embedding and other Methods}
Here, we discuss the relation between ours and other methods.
(1) \textbf{Full Embedding.} 
Both of full embedding and ours can distinguish different feature values. 
Besides, our method has the ability to reduce memory flexibly.
(2) \textbf{Hash Embedding.} 
It is a simplified form of ours, where the code block strategy is Succession, and only the first top $t$ 0-1 values are used as the embedding index.
(3) \textbf{Multi-Hash Embedding.} 
Both of them create multiple embedding indices.
But our method goes further, i.e., keeping a uniqueness constraint for these indices. 
(4) \textbf{Q-R Trick.} 
 Q-R trick is a special case of our method.
When we utilize Succession and the block number is set to 2. 
The first top $t$ 0-1 code and the left 0-1 code can be taken as the quotient and the remainder in Q-R trick respectively. 


\section{Experiments}

\noindent\textbf{Datasets.}
(1) \emph{Alibaba} is an industrial dataset which is obtained from Taobao. 
There are a total 4 billion samples, 100 million users.
(2) \emph{Amazon} \footnote{https://www.amazon.com/} is collected from the Electronics category on Amazon.
There are total 1,292,954 samples, 1,157,633 users.
(3) \emph{MovieLens} \footnote{https://grouplens.org/datasets/movielens/} is a reviews dataset and is collected from the MovieLens web site.
There are total 1,000,209 samples, 6,040 users.


\noindent\textbf{Baselines.}
(1) \emph{Full Embedding (Full)} is a standard embedding learning method.
(2) \emph{Hash Embedding (Hash)}\cite{weinberger2009feature} applies the modulo operation on the Hash ID to obtain an embedding index.
(3) \emph{Multi-Hash Embedding (MH)} \cite{tito2017hash} applies multiple hash functions to the feature value to obtain multiple indices.
(4) \emph{Q-R Trick (Q-R)} \cite{shi2020compositional} take both the remainder and the quotient as indices.



\noindent\textbf{Training Details.} 
All methods have the same EDRM architecture.
The embedding dimensionality is also set the same for all methods.
The methods (i.e., MH, Q-R trick, and ours) employ embedding table sharing strategy in different indices for memory reduction purpose and take sum pooling as the fusion function.
For MH, we use 2 hash functions as suggested by authors \cite{tito2017hash}.
For our method, the code block strategy is Succession.
We use the Adagrad optimizer  with a learning rate of 0.005.
The batch size is 1024 for all datasets.

\begin{figure}[t]
  \centering
    \subfigure[Test AUC in different epochs]{
    \includegraphics[width= .2\textwidth]{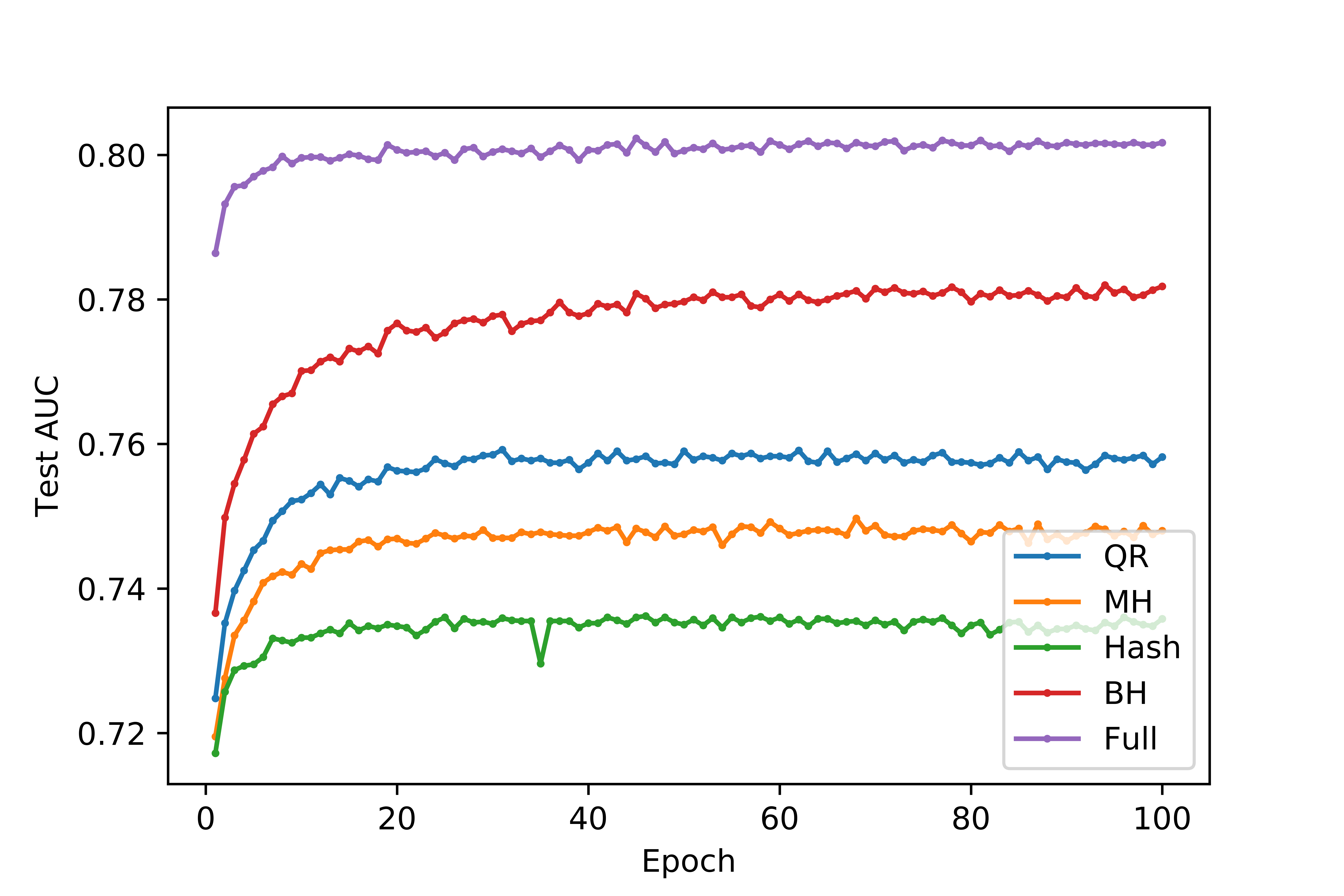}
    \label{fig:subfigure2}
  }
      \subfigure[Test loss in different epochs]{
    \includegraphics[width= .2\textwidth]{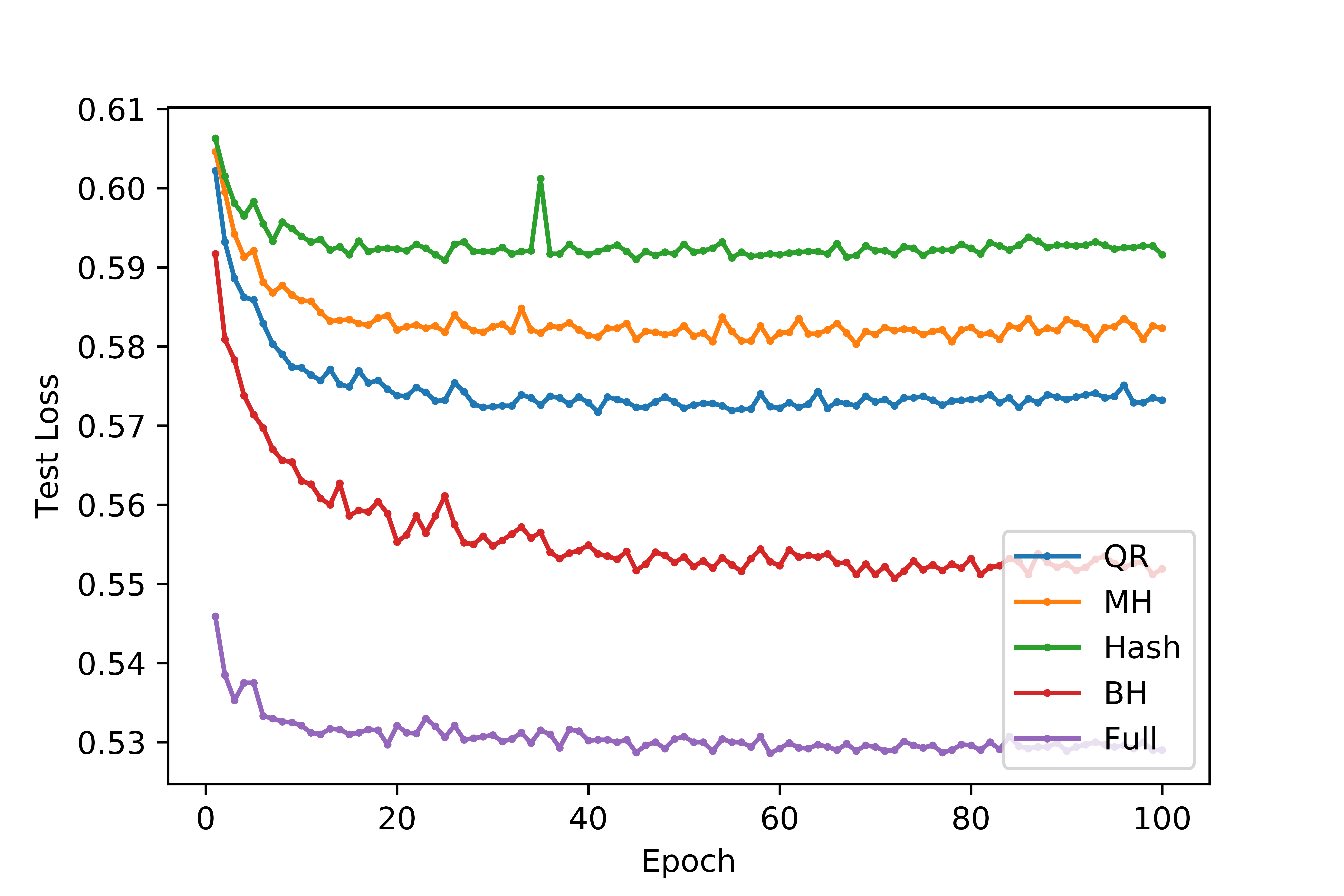}
    \label{fig:subfigure2}
  }

\vspace{-1em}
\caption{Convergence of different methods. 
}
\vspace{-1.5em}
\label{figure:Convergence of Different Models}
\end{figure}

\subsection{Click-Through Rate (CTR) Prediction Task}
We conduct experiments on CTR prediction tasks and compare the performance with different memory reduction ratios of the full embedding.
AUC (\%) \cite{fawcett2006introduction} score is reported as the metric.
Note 0.1\% absolute AUC gain is regarded as significant for the CTR task \cite{zhou2018deepdin,autoint,cheng2016wide}.
Results are shown in Table \ref{table:The AUC results for CTR tasks on Alibaba Dataset.}.
BH refers to our method. 

\noindent\textbf{Comparison with Mod-based Hash Embedding Methods.}
In general, BH performs best on all cases,
and the gain gap between BH and baselines is increased for a smaller model.
For example, compared with Q-R on MovieLens, BH can achieve 0.24\% gains when the reduction ratio is 37.5 \%.
While when the ratio becomes 0.1 \%, the gain gap is increased to 2.28 \%.
It indicates that due to the nice properties of BH (see Section \ref{sec:Desiderata}), BH can better represent each categorical feature value, especially for a tiny model. 

\noindent\textbf{Comparison with the Full Embedding.}
We can observe that: (1) Since the full embedding method contains significant parameters, it gets better performance.
(2) In most cases, BH can obtain competitive performance with the full embedding method.

\subsection{Memory Reduction Comparison}
In this section, we conduct experiments to evaluate the model size of all methods when achieving similar performance.
Specifically, we take the dataset Alibaba as an example due to its closeness with the web-scale application.
Then we report the model size of different embedding methods when they achieve 99\% performance as the full embedding method achieves in AUC score.
Besides, to further evaluate the reduction ratio, we also report the size of the embedding table of the top 3 largest for each method.
The results are shown in Table \ref{table:The results of memory size when all the methods }.
Some observations are summarized as follows:
(1) Compared with other embedding methods, when all of them archive similar scores, BH can cost the smallest memory size.
(2) Compared with the full embedding method, BH can adopt an extremely tiny model (i.e., 1000$\times$ smaller) to achieve 99 \% performance.
Such a small model with high performance is urgently needed to develop EDRMs on a memory-sensitive scenarios (e.g., mobile devices). 


\begin{table}[t] \smaller
\caption{The results of different code block strategies.}
\vspace{-1em}
\begin{tabular}{cccc}
\toprule 
         & Succession & Skip  & Q-R  \\ \hline
Alibaba & \textbf{69.90}   & 69.87  & 69.47    \\ 
Amazon & 67.59   & \textbf{67.60}  & 66.67   \\ 
MovieLens & 78.20   & \textbf{78.25}  & 75.92    \\ \bottomrule
\end{tabular}
\vspace{-1.5em}
\label{tabel:The Effect of the Different Code Block Strategy}
\end{table}

\subsection{The Effect of the Code Block Strategy}
\label{sec:The Effect of the Code Block Strategy}
In this section, we evaluate the performance of the proposed two code block strategies, i.e., Succession and Skip.
To have a fair comparison, we keep the same reduction ratio for these two strategies.
Table \ref{tabel:The Effect of the Different Code Block Strategy} shows the results.
Note we also provide the best performance (Q-R) among baselines for comparison.
We can observe that Succession and Skip achieve similar performance overall datasets, and perform better  than the best baselines.
It indicates the uniqueness of code block strategy is helpful to improve the embedding performance no matter what kind of code block strategies we choose.

\subsection{Analysis of Convergence}
We conduct experiments to analyze the convergence of different models.
Specifically, we keep the same reduction ratio for all methods and report the AUC and the loss value of these methods on test data of MovieLens within 100 epochs (similar conclusions can be found in other datasets).
As shown in Fig \ref{figure:Convergence of Different Models}, we can find:
(1) Compared the AUC and loss curves of mod-based hash embedding methods, BH converges faster than that of other baselines.
Furthermore, BH can achieve a higher AUC score and a lower loss value.
It demonstrates the effectiveness of our method when reducing EDRMs into a small-scale size.
(2) Due to more parameters adopted in full embedding, it archives the best performance.
But, BH can also achieve competitive performance compared with full embedding. 

\section{Conclusion}
In this paper, to tackle the memory problem in embedding learning, we propose a binary code based hash embedding.
A binary code is firstly generated to guarantee a unique index code.
Then a code block strategy is designed to flexibly reduce the embedding table size.
Finally, the feature embedding vector is obtained by combining the embedding vectors from different code blocks.
Experimental results show that even if the model size is 1000$\times$ smaller, we can still obtain the 99\% performance by binary code based hash embedding.



\end{document}